\def\mathcal{\cal}

\documentstyle[prl,aps,twocolumn,floats]{revtex}
\input psfig

\newcommand{\bleq}{\ifpreprintsty
                   \else
                   \end{multicols}\vspace*{-3.5ex}{\tiny
                   \noindent\begin{tabular}[t]{c|}
                   \parbox{0.493\hsize}{~} \\ \hline \end{tabular}}
                   \fi}
\newcommand{\eleq}{\ifpreprintsty
                   \else
                   {\tiny\hspace*{\fill}\begin{tabular}[t]{|c}\hline
                    \parbox{0.49\hsize}{~} \\
                    \end{tabular}}\vspace*{-2.5ex}\begin{multicols}{2}
                    \fi}
\newcommand{\bcols}{\ifpreprintsty\else\begin{multicols}{2}\fi}
\newcommand{\ecols}{\ifpreprintsty\else\end{multicols}\fi}

\def \be{\begin{equation}}
\def \ee{\end{equation}}

\def \ve{\varepsilon}

\def \tqx{\tilde{q}_x}
\def \tqy{\tilde{q}_y}
\def \tqa{\tilde{q}_{\rm a}}

\def \bPsi{ {\bf \Psi} }

\begin{document}

\bibliographystyle{simpl1}

\title{Fano resonances as a probe of phase coherence in quantum dots}

\author{A. A. Clerk, X. Waintal, P. W. Brouwer}
\address{Laboratory of Atomic and Solid State Physics,
Cornell University, Ithaca NY 14853, USA
\\
{\rm (Jan 5, 2001)}
\medskip ~~\\ \parbox{14cm}{\rm
In the presence of direct trajectories connecting source and drain
contacts, the conductance of a quantum dot may exhibit resonances
of the Fano type.  Since Fano resonances result from the
interference of two transmission pathways, their lineshape (as
described by the Fano parameter $q$) is sensitive to dephasing in
the quantum dot. We show that under certain circumstances the
dephasing time can be extracted from a measurement of $q$ for a
single resonance.  We also show that $q$ fluctuates from level to
level, and calculate its probability distribution for a chaotic
quantum dot.  Our results are relevant to recent experiments by
G\"ores {\it et al.}
}}

\maketitle


Perhaps one of the most fundamental issues in the field
of mesoscopic physics is that of phase coherence: under
what conditions are electrons able to retain a well-defined
phase? This issue is of particular interest for quantum
dots in the Coulomb blockade regime, where the
electrical conductance is suppressed except for points of
charge degeneracy \cite{Likharev}.
Despite the fact that interactions are strong
in these dots, the shape of the conductance peaks
can be well understood in terms of single-particle wavefunctions.
Unfortunately, a simple conductance measurement cannot
discriminate between coherent and incoherent (sequential)
tunneling, as both mechanisms give rise to lineshapes of
Breit-Wigner form \cite{Buttiker}. Instead, to establish phase coherence,
the quantum dot has to be embedded in an interferometer.
This was first done by Yacoby {\em et al.} \cite{Yacoby},
who included a quantum dot in one arm of an Aharonov-Bohm
ring (see also Ref.\ \onlinecite{Buks}).

\begin{figure}
\centerline{\psfig{figure=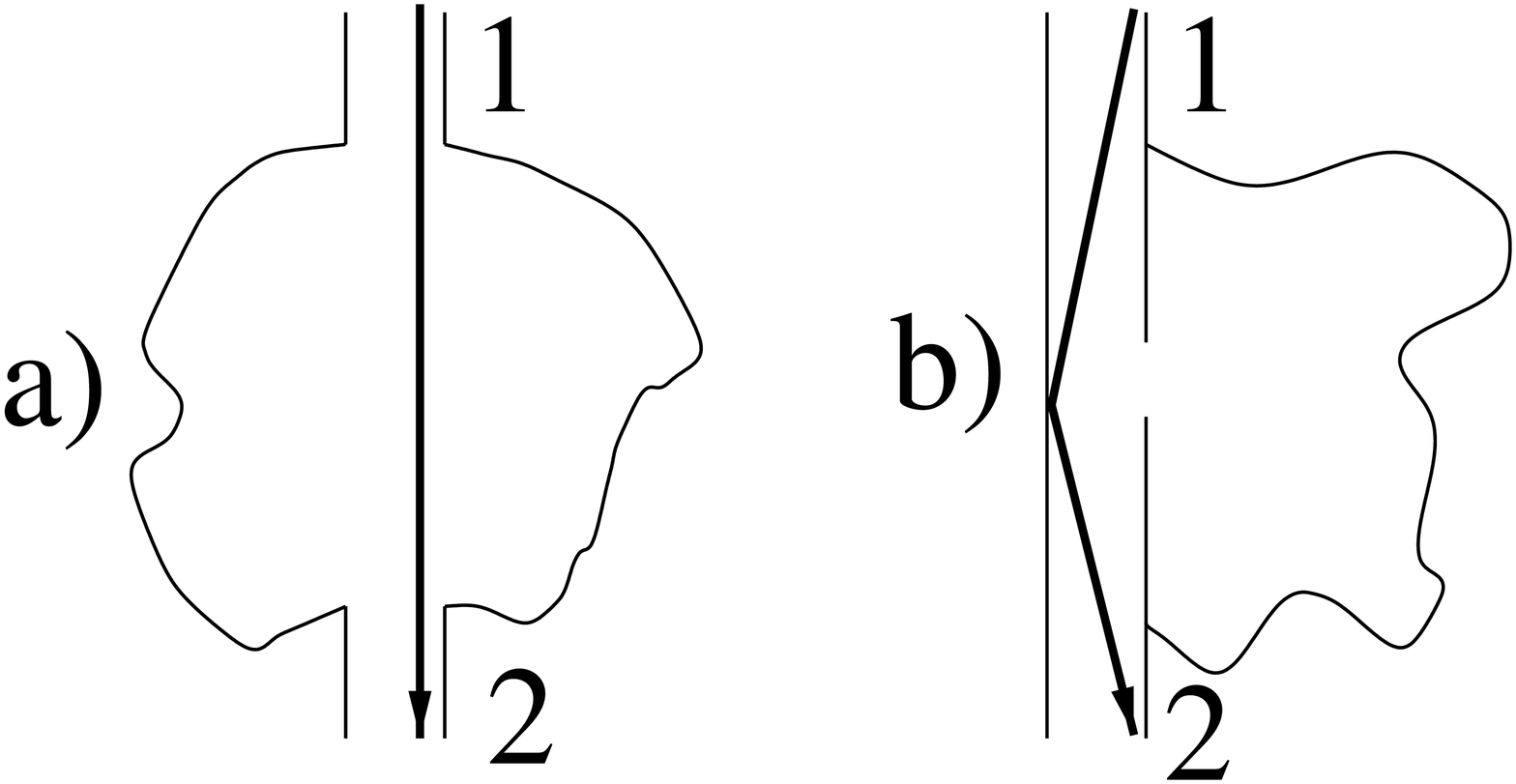,width=5.0cm}}
\refstepcounter{figure} \label{FanoDot} \label{fig:1}
\bigskip
\vspace{-10pt} { \small FIG. \ref{FanoDot}. Schematic drawing of
two quantum dot systems in which Fano resonances are expected. a)
A dot similar to that used in a recent experiment by G\"ores {\it
et al.}\cite{Gores}, showing a possible direct path ($d$). b)  A
system where the width of Fano resonances can be tuned without
altering the direct path. }
\end{figure}

Given this result, it is natural to question the {\em extent}
to which transport is phase coherent. This question could
not be fully addressed in Ref.\ \onlinecite{Yacoby}, because
dephasing in the quantum dot and in the arms of the interferometer
cannot easily be separated.  An extremely promising development
in this respect is found in a recent work by G\"ores {\em et al.}
\cite{Gores}, who observed resonances with a Fano lineshape in
the conductance of Coulomb-blockaded dots
\cite{Fano}, instead of the usual Breit-Wigner form.
Fano resonances are
caused by the interference of two transport pathways, a resonant and
a nonresonant one, and are thus sensitive to phase coherence.
In the dots of Ref.\ \onlinecite{Gores}, the direct pathway is
probably direct transmission through the dot, as schematically
depicted in Fig.\ \ref{fig:1}. In this sense, the dot serves as
its own interferometer!

Fano resonances have a lineshape of the form
\begin{equation}
  G(\varepsilon) = G_{\rm d} {|2\varepsilon + q \Gamma|^2
  \over 4 \varepsilon^2 + \Gamma^2}, \label{eq:Fano}
\end{equation}
where $G$ is the conductance, measured in units of $2e^2/h$,
$\varepsilon$ the energy, set by a gate voltage, $\Gamma$ the
resonance width, $G_{\rm d}$ the nonresonant conductance, and $q$
the (complex) ``Fano parameter''. The resonance form
(\ref{eq:Fano}) arises from the interference of a ``direct''
nonresonant path with transmission amplitude $t_{\rm d} = e^{i
\beta_{\rm d}} \sqrt{G_{\rm d}}$
and a resonant path with transmission
amplitude $t_{\rm r}(\varepsilon) = z_{\rm r}\Gamma/(2 \varepsilon
+ i \Gamma)$, where $G(\varepsilon) = |t_{\rm d} + t_{\rm r}|^2$ and
$q = i + z_{\rm r} e^{-i \beta_{\rm d}}/\sqrt{G_{\rm d}}$.
The Fano lineshape (\ref{eq:Fano}) is for temperatures $T \ll \Gamma$,
which is appropriate for the experiments on very small quantum dots
of Refs.\ \onlinecite{Yacoby,Buks,Gores}.
Examples of dots that could show Fano resonances are shown
in Fig.\ \ref{fig:1}. The example of Fig.\ \ref{fig:1}b is
particularly interesting, as it allows one to
control the width of the Fano resonances by
varying the size of the contact to the cavity.

Through the complex Fano parameter $q$, Fano resonances contain
more information than Breit-Wigner resonances. Moreover, as the
direct and resonant paths are not spatially separated, no source
of decoherence other than dephasing inside the quantum dot can
affect the lineshape.  In fact, as we will show below, in the
presence of time-reversal symmetry (TRS) or for a dot of the form of
Fig.\ \ref{fig:1}b, where the opening to the cavity contains at
most one propagating channel, measurement of a single Fano
resonance at $T \ll \Gamma$ is sufficient to determine the
dephasing time $\tau_{\phi}$ in the quantum dot: \be
\label{DephEqn} \frac{\hbar}{ \tau_\phi} = \Gamma \left( |q|^2 + 1
- \sqrt{ (|q|^2 + 1)^2 - 4 (\mbox{Im}\, q)^2}\right). \ee

In this letter, we develop a detailed description of Fano
resonances in quantum dots. In addition to the effect of
dephasing, we consider
mesoscopic fluctuations --- the Fano parameter $q$ and the
width $\Gamma$ fluctuate from resonance to resonance. We
calculate the probability distribution of $q$ for a set of consecutive
resonances in a chaotic quantum dot using random matrix theory. We close
with a comparison of our results and the experiment of Ref.\
\onlinecite{Gores}.  Several previous studies have treated Fano resonances in
quasi-one-dimensional systems rather than quantum dots \cite{OldFanos}
and without mesoscopic fluctuations or dephasing.

{\it Model}. We consider a quantum dot coupled to two single-mode
leads ($1$ and $2$) via point contacts, see Fig.\ \ref{fig:1}.
Transport through the system is characterized by the $2 \times 2$
scattering matrix $S(\varepsilon)$, which we parameterize in terms
of a (unitary) scattering matrix $S_0(\varepsilon)$ for processes
that involve ergodic exploration of the cavity, and a
(non-unitary) matrix $\bar S$ that describes scattering via the
direct, non-resonant paths (both transmitting and reflecting)
\cite{MelloReview,BeenakkerReview}, \be
\label{STotal}
  S = \bar{S} + t' \frac{1}{1 - S_0 r'} S_0 t. \ee
The auxiliary matrix $t$ describes transmission from the leads to 
an ergodic dot state, while
$t'$ describes transmission from such a state back into the 
leads.  Similarly, $r'$ describes reflection of an electron
leaving an ergodic dot state back into the dot.   
Our results are independent of $t'$, $t$, and $r'$, 
as long as the $4 \times 4$ matrix $$ \left(
\begin{array}{cc}
    \bar{S} & t' \\
    t & r'
    \end{array}   \right)
$$ is unitary.  As the time scale of the direct processes is much
smaller than $\hbar/\Delta$, where $\Delta$ is the level
spacing of the dot, $\bar S$ will be constant over an energy
interval spanning a large number of distinct resonances.  Also,
since TRS cannot be broken on this short time
scale, $\bar S$ is symmetric.  In contrast,
the scattering matrix $S_0$, which describes scattering from long,
resonant paths, depends on energy and is only symmetric
in the absence of a TRS breaking magnetic
field. We use the polar decomposition of $\bar S$
\cite{BeenakkerReview},
\begin{equation}
  \bar S = U \sqrt{(1-T)} U^{\rm T},
\end{equation}
where $U$ is a $2 \times 2$ unitary matrix and $T = \mbox{diag}\,
(T_1,T_2)$. Without loss of generality, we may choose
$t' = t^T = U \sqrt{T}$, $r' = -\sqrt{1-T}$.
The numbers $0 \le T_1 \le T_2 \le 1$ are known as
``sticking probabilities'' \cite{WeidenZuk}, i.e., the probabilities for
scattering through paths that explore the dot ergodically, instead of
direct transmission via a short path.  (The inequality $T_1 \le T_2$
is made for definiteness.) The standard theory of Fano resonances
\cite{Fano} (see also \cite{OldFanos})
assumes the existence of a single sticking probability
only. In our case, the existence of two sticking probabilities
$T_1$ and $T_2$ follows because the dot is coupled to
two single-mode leads.
We assume $T_1, T_2 \ll 1$, which ensures that the resonances are
narrow and well-separated.

For $S_0$, we make use of the
formula \cite{BeenakkerReview}
\begin{equation}
  S_0(\varepsilon)=\left[ 1-iK(\varepsilon) \right] /
    \left[ 1+iK(\varepsilon) \right],
\end{equation}
where $K$ is a $2 \times 2$ matrix representing the Green function
of the closed cavity at the contacts
\cite{LewenkopfWeidenmueller}. A resonance occurs when $K$ has a
pole, i.e., when $\varepsilon$ coincides with an energy level of
the closed dot. Close to resonance, $K \simeq \Delta \bPsi
\bPsi^{\dagger}/(\pi \varepsilon)$, where
the two-component vector $\bPsi = \left(\Psi_1,\Psi_2 \right)$ represents
the values of the wavefunction at the leads. For convenience, we
have set the resonance energy to zero. Then $S(\varepsilon)$ takes
the form \be \label{TotalS} S(\varepsilon) = U \left(
    1 -   \frac{2i \Delta \sqrt{T} \bPsi \bPsi^{\dagger} \sqrt{T}}
        {4 \pi \varepsilon + i 2 \pi \Gamma}
    \right) U^T,
\ee
where $\Gamma = \Delta \bPsi^{\dagger} T \bPsi/ 2 \pi$. By the Landauer
formula, $S(\varepsilon)$ determines the conductance
$G(\varepsilon) = |S(\varepsilon)_{12}|^2$, and hence the Fano
parameter $q$. Far from resonance, the second term in Eq.\
(\ref{TotalS}) vanishes, so that the non-resonant contribution
to the conductance reads $G_{\rm d} = |(U U^{\rm T})_{12}|^2$.

Several conclusions can be drawn
directly from Eq. (\ref{TotalS}). First, writing $q = q_x + i q_y$,
unitarity of $S$ implies that $q_x$ and $q_y$ are bounded.
Defining $(q_x^{\rm max})^2 = (q_y^{\rm max})^2 - 1 = 1/G_{\rm d} -
1$, and changing to ``normalized'' real and imaginary parts of
the Fano parameter $\tilde q_x = q_x/q_x^{\rm max}$, $\tilde
q_y = q_y/q_y^{\rm max}$, Eq.\ (\ref{TotalS})
gives the constraint
\be  \label{Elipse}
  \tilde q_x^2 + \tilde q_y^2 \le 1.
\ee
Second, in the presence of TRS, the wavefunction
$\bPsi$ can be chosen real, from which one finds $q_y = 0$. This
implies that, in the absence of dephasing, the conductance drops
to zero at $\varepsilon = -q_x \Gamma/2$.

\begin{figure}
\centerline{\psfig{figure=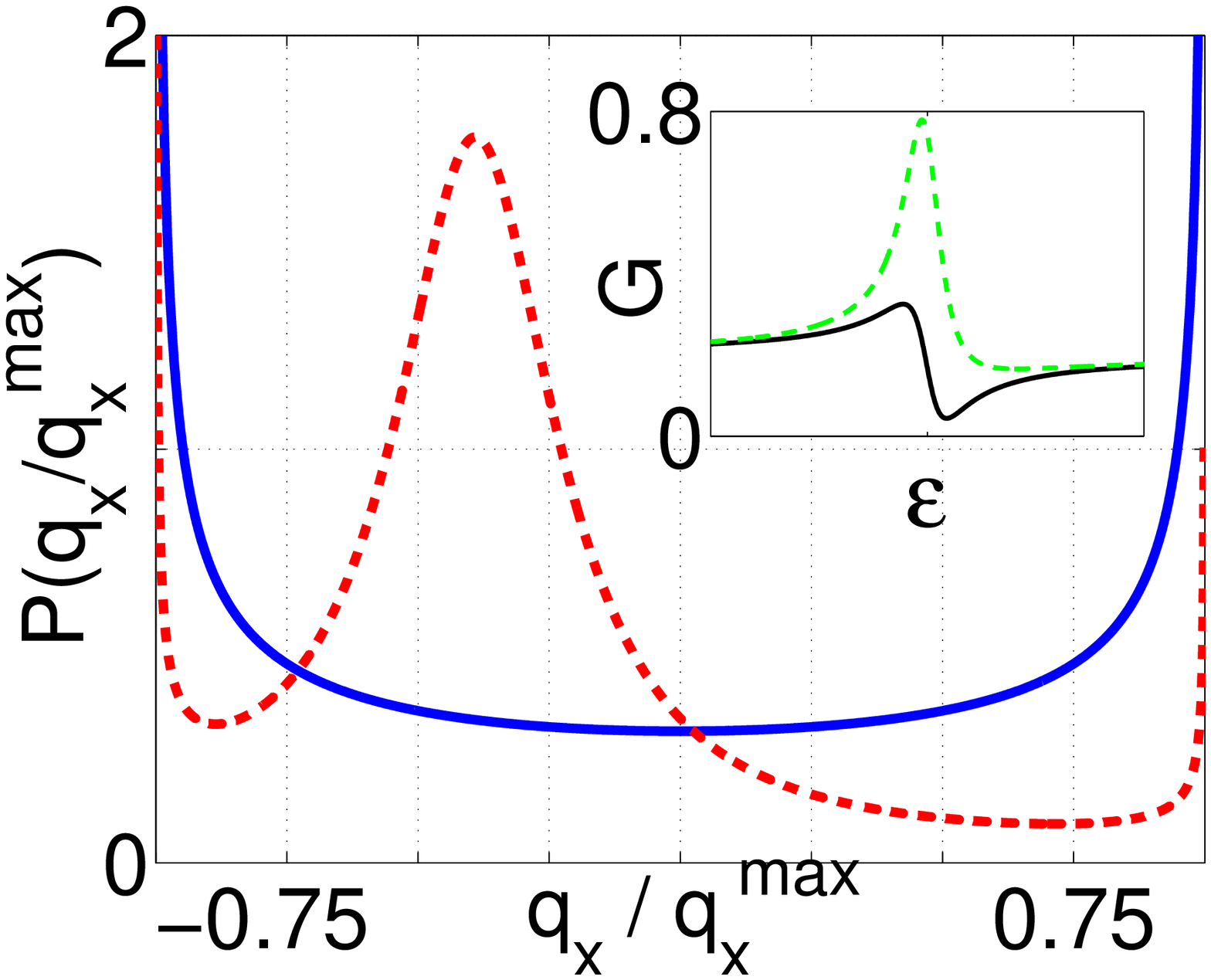,width=6.3cm}}
\refstepcounter{figure} \label{DistPlot}
\bigskip
\vspace{-10pt}
{
\small FIG. \ref{DistPlot}.
Distribution of $q_x$ in the presence of TRS
for $T_2 = T_1$ (solid line) and $T_2 = 100 T_1$ (dashed line).
Inset: Two Fano resonances with the same $q_x$, but where
$q_y$ is greater for the dashed curve compared to the solid curve.
(Breaking TRS causes $q_y$ to increase on average,
leading to more Breit-Wigner-like lineshapes.)
}
\end{figure}

{\em Resonance-to-resonance fluctuations}.  We now consider a set
of consecutive resonances in a single quantum dot, all occurring
within an energy interval in which $\bar S$ can be considered
constant.  In general, the Fano parameter $q$ is sensitive to the
resonance wavefunction $\bPsi$ only through the ratio $T_1 \Psi_1 / T_2
\Psi_2$.  If one of the sticking probabilities is zero, so that
the resonant state is only coupled to the outside world via a
single channel, this ratio is independent of $\bPsi$, and $q$
is set solely by the direct process.  This results in   
$q$ being real and the same for each resonance:
\begin{equation}
  q = q_{\rm a} =
i (U_{11} U_{21} - U_{22} U_{12})/(U_{11} U_{21} + U_{22} U_{12}).
\label{eq:qa}
\end{equation}
In quantum dots, this limit can be realized, for example, in the
geometry of Fig.\ \ref{fig:1}b, if the opening to the cavity
supports only one mode at the Fermi level. However, in the generic
case (if the opening contains more than one mode, or in the
geometry of Fig.\ \ref{fig:1}a), both sticking probabilities $T_1$
and $T_2$ are nonzero.  Then $q$ depends on the wavefunction of
the resonance, and should exhibit mesoscopic fluctuations from
resonance to resonance.  In the case of a chaotic
quantum dot, we obtain the distribution of $q$ for our set of
resonances by using random matrix theory (RMT) for the statistics
of the wavefunction $\bPsi$ \cite{Mehta}, keeping $\bar S$ the same
same for all resonances (and hence $U$ and $T$).  According to
RMT, the elements of $\bPsi$ are independently distributed real
(complex) Gaussian random numbers with zero mean and unit
variance, in the presence (absence) of TRS. In
terms of the normalized real and imaginary parts $\tilde q_x$ and
$\tilde q_y$, and with $\tilde q_{\rm a} = q_{\rm a}/q_x^{\rm
max}$, we find that in the presence of TRS the
distribution $P$ is given by
\begin{mathletters}
\label{AllDists}
\be \label{GOEDist}
P =  \frac{1}{\pi} \sqrt{\frac{1+\alpha}{1-\tqx^2}}
    \frac{1 + \frac{\alpha}{2} \left(1 - \tqx \tqa \right)}
{ 1 + \alpha \left(1 - \tqx \tqa\right) + \frac{\alpha^2}{4}
 \left(\tqx - \tqa\right)^2 }  \delta(\tqy),
\ee
while in the absence of TRS
\begin{eqnarray}
\label{GUEDist}
P &=&  \frac{1+\alpha}{2 \pi \sqrt{1-\tqx^2-\tqy^2}} \times \\
 && \frac{ \left[1 + \frac{\alpha}{2} (1 - \tqx \tqa)\right]^2
  + \frac{\alpha^2}{4}(1-\tqx^2-\tqy^2)(1-\tqa^2)}
{\Big[
    1 + \frac{\alpha^2}{4}\left[(\tqx-\tqa)^2 + \tqy^2 - \tqy^2 \tqa^2
 \right] + \alpha \left(1 - \tqx \tqa \right)
    \Big]^2}.
\nonumber
\end{eqnarray}
\end{mathletters}%
Here $\alpha = T_2/T_1 - 1 \ge 0$. (We have averaged over the
resonance width $\Gamma$, which is also a random variable
\cite{GAMMAEQN}.) In case of symmetric couplings ($T_1 = T_2$),
the distribution simplifies to $P(\tqx) = \pi^{-1}(1 -
\tqx^2)^{-1/2}$ [$P(\tqx,\tqy) = (2\pi)^{-1} (1 - \tqx^2 -
\tqy^2)^{-1/2}$] with [without] TRS. In the
extreme asymmetric regime $\alpha \gg 1$, the situation becomes
similar to the case where only a single sticking probability is
non-zero--  Eq.\ (\ref{AllDists}) tends to a delta function
distribution at $q = q_{\rm a}$, cf.\ Eq.\ (\ref{eq:qa}). Note
however that for large but finite $\alpha$, the distribution still
has an appreciable width (see Fig. \ref{DistPlot}).

{\em Dephasing}.  The effects of dephasing are treated
phenomenologically by attaching a fictitious voltage probe to
the dot \cite{Buttiker,BrouwerVP}.  This approach is not limited
to a particular microscopic mechanism, 
and can describe dephasing from both intrinsic sources (i.e. from 
electron-electron interactions) and extrinsic sources (e.g., radiation,
magnetic impurities).  However, similar to a golden rule 
calculation \cite{Imry}, it does not
account for possible interaction effects beyond lifetime
broadening which occur at low temperatures \cite{Altshuler}.  
In practice, we first replace
$\varepsilon \to \varepsilon + i \hbar/2 \tau_{\phi}$ in Eq.\
(\ref{TotalS}), where $\tau_{\phi}$ is the phenomenological
dephasing time.  The imaginary part of $\ve$ models escape through
the fictitious voltage probe.  A
correction term is then added to the conductance formula to account
for the incoherent injection of electrons from the voltage probe
\cite{Buttiker,BrouwerVP},
\begin{equation}
  G(\varepsilon) =
  |S_{12}|^2  +
  {(1 - (S S^{\dagger})_{11}) (1 - (S S^{\dagger})_{22}) \over
    2 - (S S^{\dagger})_{11} - (S S^{\dagger})_{22}}.
\end{equation}
The second term corresponds to incoherent transmission through the
dot, and has a Breit-Wigner lineshape.  As a result, the imaginary
part $q_y$ of the Fano parameter is increased. Inclusion of
dephasing also changes the resonance width $\Gamma$ to $\Gamma +
\hbar/(2 \tau_{\phi})$. Writing the ratio of resonance widths
without and with dephasing as $\chi_{\phi} = \Gamma/(\Gamma +
\hbar/(2 \tau_{\phi}))$, we find that the change of the Fano
parameters upon inclusion of dephasing is given by:
\begin{mathletters}
\label{DephGroup}
\begin{eqnarray}
  q_x & \to & \chi_{\phi} q_x, \\
  (q_y)^2 & \to &  1 - \chi_{\phi} + \chi_{\phi}(q_x^2 + q_y^2 -
  \chi_{\phi} q_x^2). \label{eq:qxy}
\end{eqnarray}
\end{mathletters}%
In the presence of TRS, or in the extreme
asymmetric limit, where only one sticking probability is nonzero,
$q_y = 0$ in the absence of dephasing. Hence, measurement of a
nonzero $q_y$ in those cases can be used to determine 
$\tau_{\phi}$ \cite{Experiment}.
Calculating $\tau_{\phi}$ from Eq. (\ref{DephGroup}) with $q_y = 0$
in the absence of dephasing yields the relationship (\ref{DephEqn}),
as advertised. If TRS is broken and if both
sticking probabilities are finite, $q_y$ is already nonzero in
the absence of dephasing, and a measurement of $q_y$ cannot be
used to find $\tau_{\phi}$. Note that as $\tau_{\phi} \to 0$,
$G(\ve) \to G_D$, consistent with earlier work on resonant tunneling
\cite{StoneLee}.

{\it Role of Coulomb interactions}.  So far we have not addressed
the issue of Coulomb interactions, which are certainly present and
important for small quantum dots. In this respect, we note that
the time needed to traverse the quantum dot via a direct
trajectory is of the same order or smaller than the inverse
charging energy $E_c$. Hence, by the time-energy uncertainty
principle, transmission via direct paths is not forbidden by
Coulomb blockade. In fact, as was shown by Matveev and coworkers
\cite{Matveev,FM95a}, Coulomb interactions actually {\em enhance}
the probability of direct processes --- both direct reflection and
direct transmission \cite{BrouwerAleiner} --- at the cost of
ergodic scattering.  Hence, by virtue of Coulomb interactions, the
dot is driven to the weak coupling regime $T_1, T_2 \ll 1$. Such
an interaction-induced renormalization of the coupling between the
dot and its environment may explain why in the the experiment of
Ref.\ \onlinecite{Gores} sharp resonances were observed, despite
the presence of diffraction at the point contacts.  To apply our
theory to this situation, it is necessary to assume that the
renormalization of the scattering parameters for direct processes
has already taken place.  We also implicitly assumed that
interactions play no further role in modifying the resonances, and
that a single-particle approach is thus valid close to resonance,
see Refs.\ \onlinecite{Hacken,OregGefen}.

Two interesting observations of the experiment \onlinecite{Gores}
can be interpreted with the results of this letter. First, in
Ref.\ \onlinecite{Gores}, it was seen that application of a
magnetic field tended to make resonances more
Breit-Wigner like. Our calculation indicates that, for the
generic case when both sticking probabilities are nonzero,
breaking of TRS generically leads to an
increase of $q_y$, and thus to more Breit-Wigner-like resonances,
cf. Eq. (\ref{GUEDist})
(in the absence of dephasing $q_y$ was zero without a magnetic
field.) This is illustrated in the inset of Fig.\ \ref{DistPlot}
(compare against Fig.\ 6 of Ref. \cite{Gores}).

\begin{figure}
\centerline{\psfig{figure=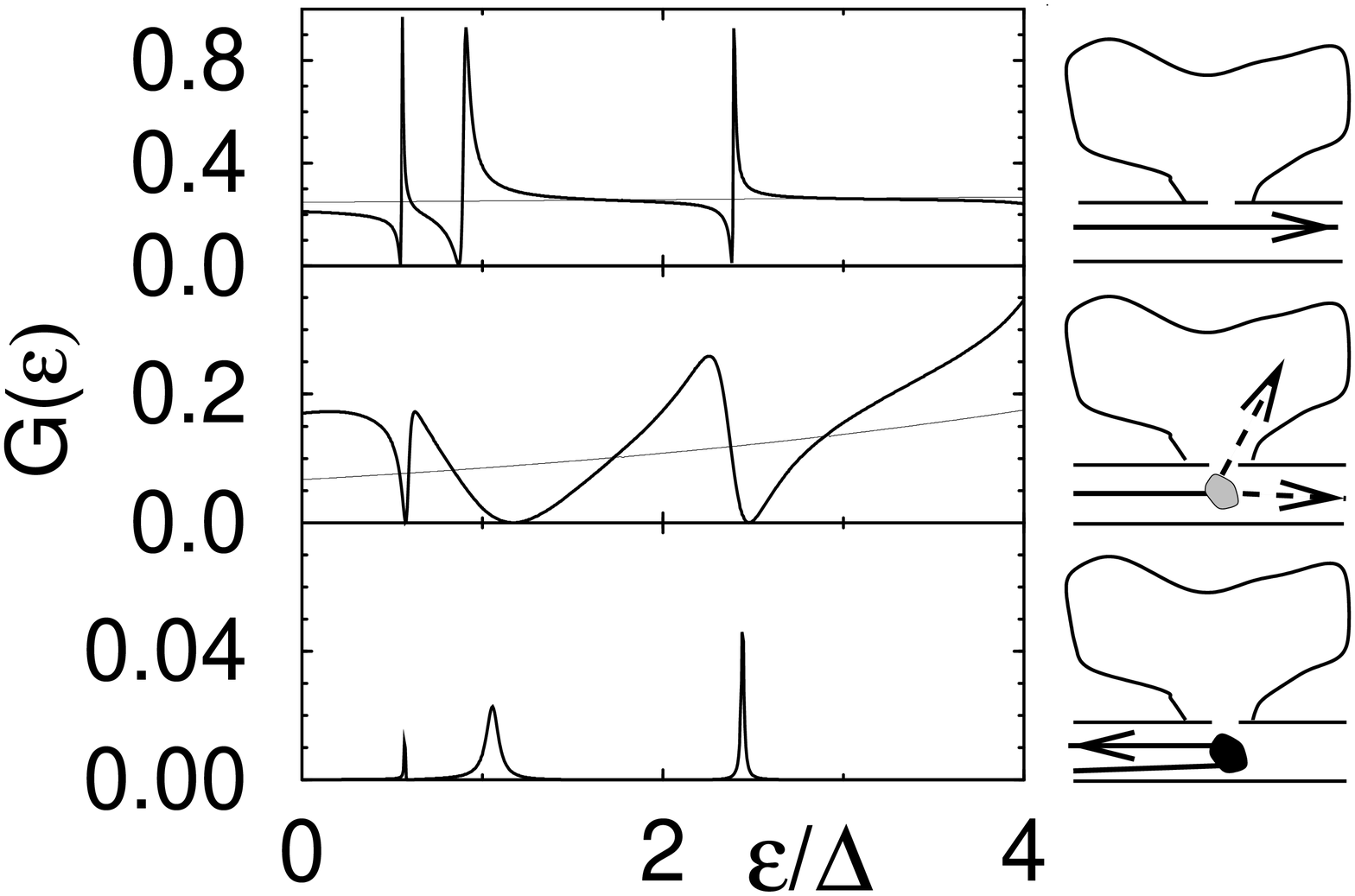,width=6.8cm}}
\refstepcounter{figure} \label{SimulFig}
\bigskip
\vspace{-40pt} { \small FIG. \ref{SimulFig}. Numerical simulation
of conductance vs. voltage showing Fano resonances for a quantum
dot with direct transmission (right). The thin line corresponds to
the background conductance (i.e., when the dot is closed off). The
three plots correspond to the same resonances for increasing value
of the scattering strength of an impurity near the contact. }
\end{figure}

Another observation in Ref.\ \onlinecite{Gores} was
that, as a function of the gate voltage that controls
the transparencies of the point contacts, the width of the observed
resonances was non-monotonic.  The conductance peaks started as
narrow Breit-Wigner resonances when the dot was pinched off ($G_d = 0$),
then
widened as the contacts were opened into
resonances exhibiting the Kondo effect.  As the contacts
were opened further, the resonances became more narrow
and had the Fano form with background conductance $G_d \simeq e^2/h$.
One possible explanation is that
diffraction at the contacts to the dot is strongest at
intermediate point contact transparencies, leading to large
sticking probabilities.
(A schematic picture of the dot
of Ref.\ \onlinecite{Gores} is shown in Fig.\ \ref{fig:1}a.) To
test such a scenario, we have performed numerical simulations of the
system shown in Fig. \ref{FanoDot}b, using a recursive Greens function
algorithm \cite{Jalabert}.
As expected, we find conductance resonances with a Fano
lineshape; typical results are shown in
Fig. \ref{SimulFig}a.
In order to simulate how the non-monotonic resonance width in
the experiment might happen, we have placed an impurity
near the opening of the dot (see Fig.\ref{SimulFig}),
and varied its scattering strength
$V$ ($V$ is the potential of the impurity sites in the simulation).
Resonances for two values of $V$ are shown
in Figs.\ \ref{SimulFig}b
and c; the resonance width $\Gamma$ does indeed exhibit a
non-monotonic dependence on $V$.
Initially increasing $V$ from zero has the effect of deflecting
more electrons into the dot and hence increasing $\Gamma$; larger
values of $V$, however, cause electrons to backscatter away from
the dot altogether, thus reducing $\Gamma$ and suppressing the
background conductance $G_{\rm d}$.

{\em Conclusion.} 
Fano resonances provide a powerful tool
for the study of phase coherence in transmission through quantum
dots.  We have shown that in certain cases,
measurement of a single resonance already allows for the
determination of $\tau_\phi$.  We have also
calculated the distribution of Fano parameters for a
chaotic quantum dot.

We thank V. Ambegaokar, D. Goldhaber-Gordon, M.
Kastner, and C. Marcus for useful discussions. A.C. acknowledges
support of the Cornell Center for Materials Research.



\begin{references}
\bibitem{GAMMAEQN}
    The distribution of widths $\Gamma$ after averaging over
    $q$ and in the presence of TRS is given by
    $P = (2 \Gamma_1 \Gamma_2)^{-1/2} 
	e^{-\Gamma/4 \Gamma_1} e^{-\Gamma/4 \Gamma_2}
    I_0 
    \left(\frac{\Gamma |\Gamma_1 - \Gamma_2|}{4 \Gamma_1 \Gamma_2} \right)$,
    where $I_0$ is a modified Bessel function, and $\Gamma_j
    = \frac{\Delta}{2 \pi} T_j$.  In the absence of TRS, one has
    $P = (\Gamma_1 - \Gamma_2)^{-1} \left(
	e^{\Gamma / \Gamma_1} - e^{\Gamma / \Gamma_2}
	\right)$.
\end{references}
\end{document}